\documentclass[11pt]{article}

\usepackage[preprint]{acl}

\usepackage{times}
\usepackage{latexsym}
\usepackage{booktabs}
\usepackage{tabularx}
\usepackage{amsmath}
\usepackage{makecell}
\usepackage{longtable}
\usepackage{booktabs}
\usepackage{xcolor}
\usepackage[T1]{fontenc}

\usepackage[utf8]{inputenc}

\usepackage{microtype}

\usepackage{inconsolata}

\usepackage{graphicx}

\usepackage{booktabs}     
\usepackage{multirow}     
\usepackage{makecell}     
\usepackage{array}        
\usepackage[table]{xcolor}
\definecolor{qred}{HTML}{E4002B}   
\definecolor{lightred}{HTML}{FDE9EC}
\definecolor{hiGreen}{HTML}{C6EFCE} 

\title{Spatial Audio Question Answering and Reasoning on Dynamic Source Movements}

\author{Arvind Krishna Sridhar\\
  Qualcomm Technologies Inc. \\
  San Diego, CA  \\
  \texttt{arvisrid@qti.qualcomm.com} \\\And
  Yinyi Guo \\
  Qualcomm Technologies Inc. \\
  San Diego, CA  \\
  \texttt{yinyig@qti.qualcomm.com}  \\\And
  Erik Visser \\
  Qualcomm Technologies Inc. \\
  San Diego, CA  \\
  \texttt{evisser@qti.qualcomm.com} \\}

\begin{document}
\maketitle
\begin{abstract}
Spatial audio understanding aims to enable machines to interpret complex auditory scenes, particularly when sound sources move over time. In this work, we study Spatial Audio Question Answering (Spatial AQA) with a focus on movement reasoning, where a model must infer object motion, position and directional changes directly from stereo audio. First, we introduce a movement‑centric spatial audio augmentation framework that synthesizes diverse motion patterns from isolated mono audio events, enabling controlled and scalable training data generation. Second, we propose an end‑to‑end multimodal finetuning approach with a thinking mode, which allows audio‑language models to produce explicit intermediate reasoning steps before predicting an answer. Third, we investigate the impact of query‑conditioned source separation as a preprocessing stage and compare three inference regimes: no masking, an audio grounding model (AGM), and ground‑truth masks. Our results show that reasoning amplifies the benefits of source separation, with thinking mode showing significant improvement of +5.1\% when a single event is present in the question. These findings highlight the interplay between movement modeling, reasoning, and separation quality, offering new insights for advancing spatial audio understanding.
\end{abstract}

\section{Introduction}
Recent advances in audio question answering (AQA) have led to the development of state-of-the-art models capable of generating relevant answers from a given question and an audio clip. Building on this progress, we believe that spatial audio question answering (SAQA) represents the next frontier in context-aware audio understanding. SAQA extends AQA to multichannel audio, implicitly integrating several traditional audio understanding tasks, including audio event detection (AED), audio captioning, audio localization and detection \cite{shimada2025stereo}, source separation \cite{nugraha2016multichannel}, and object tracking \cite{li2025patch}. Spatial AQA has applications in smart surveillance, robotics, autonomous vehicles, and augmented reality. To interpret complex auditory scenes in these domains, SAQA systems must recognize diverse audio events and reason about dynamic source movements. This work advances SAQA with a focus on movement reasoning, bridging the gap between static spatial representations and real-world dynamics. The DCASE challenge Sound Event Localization and Detection (SELD) task \cite{shimada2025stereo} has significantly advanced research in spatial audio by promoting models that jointly estimate event classes and their spatial attributes. \cite{sudarsanamtowards} propose a spatial audio question answering model that predicts audio events, temporal order of events and spatial attributes ordering and spatial positions as a classification task. Although these models have achieved strong performance on benchmark datasets, they do not support open ended question answering and reasoning on top of the attributes. We develop an end-to-end spatial audio question answering model that includes a query conditioned audio source separation with the help of an Audio Grounding Model (AGM) to reduce the complexity of the audio depending on the question. Text-to-audio grounding is the task of detecting audio events in an audio clip based on natural language \cite{yasuda2022echo}. From here on, we refer to this model as the Audio Grounding Model (AGM) for brevity. Similar to CLAP \cite{yuan2024tclaptemporalenhancedcontrastivelanguageaudio}, ELSA \cite{spatially-aware} consists of a spatial audio encoder and a text encoder that learns a joint embedding of spatial attributes and semantic information via contrastive learning. Although the spatial audio representation learned by ELSA is well-suited for tasks like semantic retrieval and spatial audio captioning, it does not perform explicit question answering and lacks fine-grained temporal reasoning.

\begin{figure*}[t!]
\centering
\begin{minipage}[b]{1\linewidth}
  \centering
  \centerline{
  \includegraphics[width=1\textwidth,scale=1.0]{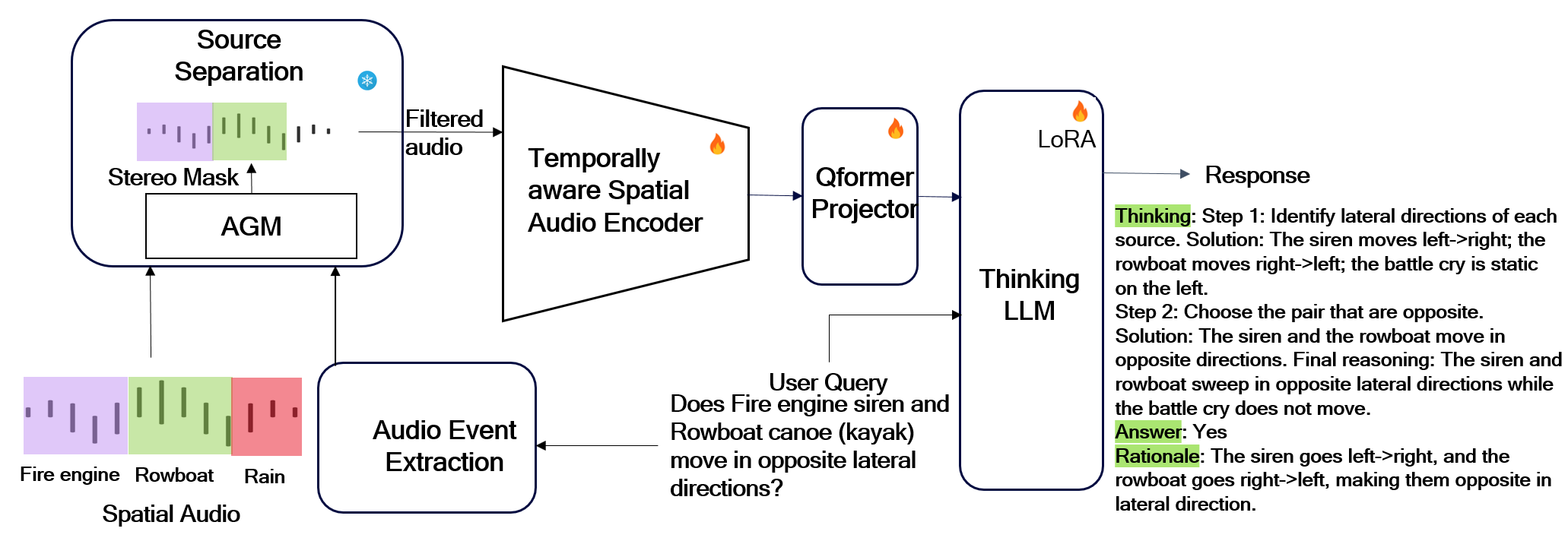}
  }
\end{minipage}
\caption{Our proposed spatial audio language model architecture with thinking mode. AGM refers to audio grounding model.}
\label{fig:Spatial LALM}
\end{figure*}

BAT \cite{zheng2024bat} introduced the first end-to-end SAQA model, combining a spatial audio encoder with a large language model (LLM) to process binaural audio and project it into a text-aligned representation for question answering. However, BAT supports only static sources and does not provide reasoning for it's answers. \cite{sakshi2025spur} develop a plug and play spatial adapter that takes FOA input and is finetuned on top of a frozen LALM to introduce general spatial perception. During the time of this writing, we don't have access to their model implementation. Therefore, we only compare our model performance against BAT as a baseline.

To the best of our knowledge, we are the first to make the following contributions:
\begin{itemize}
\item \textbf{Movement-centric Spatial Audio Dataset Augmentation.}
We introduce a dataset generation pipeline that creates spatial audio clips and corresponding question–answer pairs from the AudioSet strong‑labeled corpus. We isolate high‑quality mono audio events, filter them using quality heuristics, and introduce controlled motion by assigning predefined trajectories before spatializing them into stereo scenes.
\item \textbf{End‑to‑end spatial audio reasoning with Thinking mode.}
We build an end‑to‑end spatial audio-language model that uses a Thinking mode to produce intermediate reasoning steps before answering. This improves both interpretability and performance over non‑thinking variants. As shown in the Figure \ref{fig:Spatial LALM}, the end-to-end system comprises of a temporally aware spatial audio encoder, a Q‑former projector \cite{zhang2023vision}, and a language model and is trained jointly with component‑specific learning rates.
\item \textbf{Query‑conditioned source separation for Spatial AQA.}
We evaluate source separation as a preprocessing step and show that extracting query‑relevant audio sources leads to better attention allocation and improved accuracy. We compare no masking, an Audio Grounding Model (AGM), and ground truth masks, and demonstrate that separation synergizes strongly with thinking mode. 
\end{itemize}


\section{Related Work}
\subsection{Audio Language Models with Reasoning}
Audio Language Models (ALMs) such as Mellow \cite{deshmukh2025mellow}, GAMA \cite{ghosh-etal-2024-gama}, and Audio Flamingo 3 \cite{goel2025audioflamingo3} commonly follow a three-stage design that includes an audio encoder, an audio-to-text projection module, and a large language model. Training these components jointly has been shown to improve audio-text alignment, which in turn enhances downstream audio reasoning capabilities. The survey by \cite{cui2024survey_speechLMs} provides a broader perspective on the development of audio and speech language models, including architectural trends, pretraining strategies, and evaluation methods. \cite{sridhar2025enhancing} demonstrate that a curriculum-based fine-tuning approach can improve a model’s ability to capture temporal structure and event order without degrading its overall performance on general audio tasks. Audio-CoT \cite{ma2025audio_cot} explored chain-of-thought reasoning with self-consistency for LALMs across music, speech, and sound, and demonstrated performance improvements on the MMAU benchmark \cite{sakshi2024mmau}. Audio-Reasoner \cite{zhifei-etal-2025-audio} introduced structured reasoning involving planning, caption grounding, step-by-step reasoning, and summarization. Inspired by these developments, our work introduces a structured reasoning framework specifically designed for spatial audio source-movement understanding. Our method employs a three-stage process, namely, step-by-step planning and solution derivation, answer generation, and rationale articulation, to enhance interpretability and robustness in spatial audio reasoning tasks.

\subsection{Spatial Audio Understanding}
\cite{tang2024can} develop a spatial audio language model operating on first-order ambisonics (FOA) to assess its capability for localization, far-field recognition, and spatially informed extraction on the Spatial LibriSpeech dataset. \cite{you2026worldmonoenablingspatial} propose a unified hierarchical auditory scene analysis framework composed of a semantic audio understanding branch and a spatial processing branch, enabling LALMs to incorporate explicit spatial reasoning mechanisms. \cite{biswas2025owlgeometryawarespatialreasoning} present OWL, which uses a geometry-aware spatial audio encoder trained with room impulse response (RIR) prediction as an auxiliary task while requiring only audio at inference time. Building on these works, our paper investigates spatial audio question answering with an explicit focus on movement reasoning.

\section{Methodology}
Our methodology has three components. First, we outline how spatial audio data are generated by filtering isolated mono events, assigning motion trajectories, and spatializing them into stereo scenes paired with questions and answers. Second, we describe the multimodal spatial reasoning model. Third, we detail how query‑conditioned source separation is applied using an Audio Grounding Model (AGM) to mask stereo channels and isolate audio relevant to each query.

\begin{figure*}[t!]
\centering
\begin{minipage}[b]{1\linewidth}
  \centering
  \centerline{
  \includegraphics[width=1\textwidth,scale=1.0]{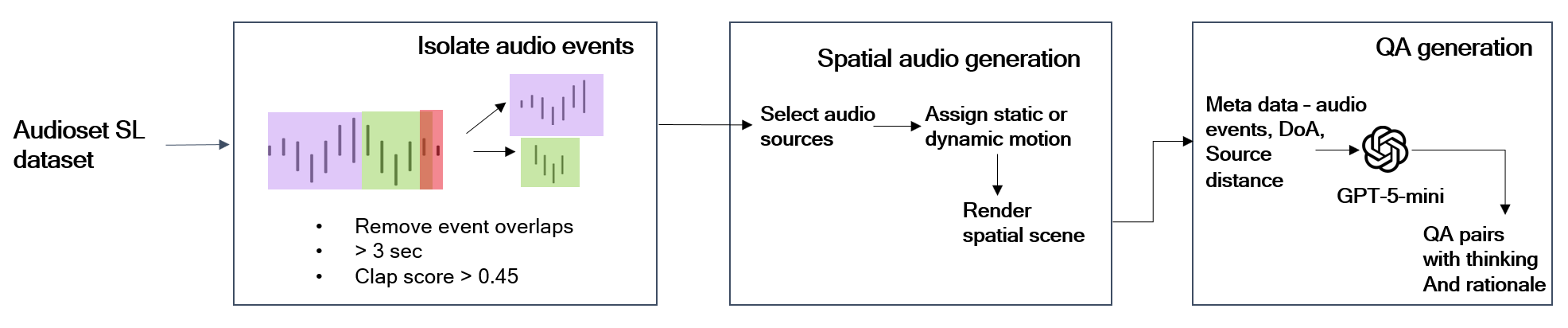}
  }
\end{minipage}
\caption{Spatial audio dataset generation}
\label{fig:spatial audio dataset generation}
\end{figure*}

\subsection{Movement\textendash Centric Spatial Audio Dataset Augmentation}
\label{sec:dataset-augmentation}

To study spatial audio reasoning with moving sound sources, we construct a controlled dataset of spatialized audio clips paired with movement\textendash focused question--answer (Q/A) annotations.
Our pipeline has three major steps: (1) isolating audio events and quality filtering (2) spatial scene synthesis with controlled motion, and (3) question-answer pair generation for movement reasoning.

\subsubsection{Isolating Audio Events and Quality Filtering}
\label{sec:event-isolation}

We begin by extracting isolated audio events from the \emph{AudioSet Strong\textendash labeled} subset, which provides frame\textendash level annotations indicating when specific events occur. We apply two quality filters when selecting isolated audio events. 
First, each event must have a minimum duration of at least 3\,s to ensure that the segment is long enough to support motion synthesis. 
Second, we require a CLAP audio–text similarity score of at least 0.45 between the event’s audio embedding and the caption 
\emph{``a sound of \textless event name\textgreater''}, which helps remove mislabeled or low‑fidelity segments.
After filtering, we obtain \textbf{224} high\textendash quality mono audio events suitable for spatialization.

\subsubsection{Spatial Scene Synthesis With Controlled Motion}
\label{sec:scene-synthesis}

Using the filtered events, we synthesize 10 second stereo spatial audio scenes.
Because our goal is movement\textendash based spatial reasoning with a stereo microphones, sources are positioned only in front of the listener, with azimuth angles sampled from $[-90^{\circ},\, 90^{\circ}]$ where $0^{\circ}$ is straight ahead, $-90^{\circ}$ is left and $90^{\circ}$ is right.
Source distance ranges from $[0,\,6]$ meters.

Each scene contains up to three events, and each event is randomly assigned either a static or a dynamic motion pattern. 
For dynamic sources, we allow several types of trajectories commonly used in spatial movement reasoning: 
\textbf{approach} motion, where the source moves closer to the listener (decreasing distance); 
\textbf{recede} motion, where the source moves away (increasing distance); 
\textbf{lateral movement}, involving left-to-right or right-to-left motion; 
and \textbf{arc trajectories}, where the source follows a curved path around the listener at a fixed radius, resulting in azimuth changes over time while the distance remains constant.

We render all scenes with Pyroomacoustics \cite{8461310}, using a maximum reflection order of $2$ and surface absorption coefficient of $0.25$, introducing modest reverberation while preserving salient spatial cues.

\subsubsection{Question\textendash Answer Pair Generation for Movement Reasoning}
\label{sec:qa-generation}

In the second stage of our pipeline, we construct question--answer (Q/A) pairs that evaluate movement understanding in Spatial AQA. For each synthesized audio clip, we store detailed metadata that includes framewise event identities, the per-frame azimuth and distance trajectories of each source, temporal boundaries and ordering such as event onsets and offsets, and a compact summary describing the overall movement patterns present in the scene. This metadata enables us to generate questions that require reasoning about spatio-temporal relationships rather than simply predicting numeric values.

While prior works such as BAT \cite{zheng2024bat} focuses on QA pairs that require direct estimation of DoA and distance, real-world queries often involve understanding spatial-temporal audio event relationships rather than purely numerical. We focus on generating diverse questions across 7 categories as shown in the Table \ref{Table:Example QA pairs} including trajectories, radial changes, comparative diagnostics, temporal sequence, relative motion between events and natural perception.

To generate questions automatically, we provide \texttt{GPT-5-mini} \cite{gpt5mini} with a structured scene description that includes the event list, azimuth and distance trajectories, onset/offset times, and a short summary of overall motion. To enable robustness to different question formats, the model is prompted to produce Boolean, multiple-choice, and open-ended questions. Apart from QA pairs, we also generate rationale to aid in interpretablity and thinking steps to improve the accuracy of the predicted answers. Depending on the difficulty of the question, we vary the thinking steps from 1 to 3. For more details on the QA data generation, please refer to the prompts in Appendix \ref{app:spatial-audio-qa-prompt}.

We conducted human evaluation of the synthesized spatial QA pairs to asses the quality of the dataset. As shown in Table \ref{tab:human-eval-overall}, the annotated scores are high, indicating that the generated questions are answerable, reasoning chains are coherent and the queries remain well grounded in their corresponding spatial audio scenes. 

We evaluate the boolean and multiple choice questions via keyword match. We utilize \texttt{GPT-5-mini} \cite{gpt5mini} for evaluating thinking steps, rationale and open ended reasoning. For evaluating thinking steps, we require the LLM to score across the following dimensions: logical coherence, step completeness, factual accuracy and alignment with ground truth. We evaluate the rationale on conciseness, accuracy and clarity. Open ended questions are evaluated on factual accuracy and semantic similarity. For more details on the prompts, please refer to Appendix \ref{app:spatial-audio-qa-evaluation-prompt}.

\begin{table}[t]
\centering
\small
\setlength{\tabcolsep}{5pt}
\caption{Human evaluation of the Spatial AQA dataset (1-5 scale; higher is better) across 200 randomly selected samples.}
\label{tab:human-eval-overall}
\begin{tabular}{l c}
\toprule
\textbf{Metric} & \textbf{Score} \\
\midrule
Answer Accuracy        & $4.93$ \\
Reasoning Quality      & $4.94$ \\
Groundedness           & $4.97$ \\
\bottomrule
\end{tabular}
\end{table}

\begin{table*}[ht]
\centering
\caption{Example QA pairs by category}
\label{Table:Example QA pairs}
\renewcommand{\arraystretch}{1.2}
\begin{tabular}{|p{2cm}|p{2.5cm}|p{5cm}|p{6cm}|}
\hline
\textbf{Category} & \textbf{Question Type} & \textbf{Example Question} & \textbf{Answer} \\
\hline
Trajectories & yes\_no & Is the Croak sound moving from left to right? & Yes \\
\hline
Radial changes & yes\_no & Are the waves moving closer to the listener as they sweep across? & Yes \\
\hline
Comparative diagnostics & multiple\_choice & Which source sweeps across a wider left$\rightarrow$right range? \newline \emph{Options: [Blender, Car, Both equally, Neither]} & Blender \\
\hline
Temporal sequence & open & Narrate the overall choreography: where does each source start, how does it move, and where does it finish? & The whale begins on the left and moves a moderate amount further left while approaching the listener; the lawn mower starts on the right, makes a wide sweep across the center to the left while coming closer and then moving away, and finishes on the left — resulting in the lawn mower crossing through the middle and the whale remaining on the left throughout. \\
\hline
Relative motion between events & yes\_no & Do scissors and cattle sound move toward the same region (converge) as they progress? & Yes \\
\hline
Natural perception & open & Which source feels most noticeable near the middle of the scene and why? & The Applause feels most noticeable near the middle because it crosses the center with a moderate sweep, passing through that region, whereas the Camera stays on the left and does not pass through the middle. \\
\bottomrule
\end{tabular}
\end{table*}

\subsection{Multimodal end-to-end training:}
Our end-to-end Spatial Audio Language Model consists of three key components as show in the Figure \ref{fig:Spatial LALM}: a temporally aware spatial audio encoder, a Q-Former projection module \cite{zhang2023vision}, and a large language model (LLM) with thinking capabilities finetuned with LoRA. The Spatial audio encoder takes stereo audio as input and encodes them into audio tokens. We adopt the spatial audio encoder from BAT with a modification to it's penultimate layer to incorporate temporal awareness enabling frame-level reasoning over audio events and spatial attributes. 
We utilize a Q-Former to bridge the acoustic feature embedding from spatial audio encoder to a Thinking LLM (Qwen 3 4B) \cite{qwen3technicalreport}, enabling the model to generate internal thinking steps before arriving at the answer and it's rationale. 

To enable robust spatial movement reasoning, we employ a variable learning framework that helps in learning the spatial audio to text. In the first stage, Foundational spatial grounding, the model is trained on spatial audioclips with frozen spatial audio encoder to allow the model to learn spatial audio to text mappings by learning the projection module weights from scratch. This stage helps the model in learning a clean mapping between the audio source signals and their corresponding trajectories in audio-text semantic space. 

In the second stage, we train all the three components including the spatial audio encoder, Qformer projection and the thinking LLM. We introduce realistic environment constraints such as max order 2 room reflections and allows overlap of audio events upto 30\% requireing the model to disentangle concurrent spatial trajectories. To ensure stable gradient updates and avoid catastrophic forgetting of clean signal from stage 1, we implement a balanced mini-batch sampling strategy with 1:1 ratio of audio clips with overlapping and non-overlapping audio events.  

\section{Audio source separation}
When answering a question, humans naturally focus only on the few audio events that are relevant to the query rather than the entire soundscape. To mimic this behavior, we extract the key audio events mentioned in the question through keyword matching and pass this filtered set of event labels to the source‑separation module, which then isolates only the query‑relevant audio from the spatial clip.

To support flexible on-device deployment across diverse acoustic conditions and use cases, we develop an open vocabulary Sound Event Detection (SED) model that grounds free form text queries to corresponding audio events. Rather than relying on a fixed label set, the model localizes sound events conditioned on arbitrary textual descriptions of interest. Our Audio Grounding Model (AGM) is trained following the phrase level WSTAG framework in \cite{xu2024weaklysupervisedtexttoaudiogrounding} using the AudioCaps \cite{kim-etal-2019-audiocaps} dataset, from which we derive phrase level supervision. AGM adopts the same architecture and hyperparameters as \cite{xu2024weaklysupervisedtexttoaudiogrounding}: a CRNN based audio encoder with eight convolutional layers followed by a bidirectional GRU, and a text encoder consisting of a word embedding layer with mean pooling. Frame level grounding scores are computed via cosine similarity between audio frame embeddings and text tag embeddings, followed by a sigmoid activation. To extract temporal event boundaries, we apply a global threshold of 0.8 to the framewise scores and refine them with a 0.3 second median filter. These boundaries are then used to mask the input audio by zeroing out waveform regions outside the detected event span; if no boundary is detected, no masking is applied. We utilize AGM as an inference time pre-processing step for filtering relevant audio events. In future, we plan to train the audio separation module as well during the end-to-end training.

\section{Experiments}
\subsection{Datasets and Models}
\label{sec:datasets}
We choose the BAT model as our baseline. We use Audioset strong labelled to construct the spatial audio clips using pyroomacoustics. We generate spatial audio captions and question answer pairs using LLM. Our main architecture consists of spatial audio encoder from BAT with a modification in the penultimate layer to support temporal awaress across frames, Qformer \cite{zhang2023vision} as projection module and Qwen 3 4B \cite{qwen3technicalreport} thinking as the LLM. We use source separation as a processing step during inference. We require the model to generate answer and a rationale for interpretability.

\subsection{Experiment Setup}
We train the AQA architecture from scratch with end-2-end learning as described in the methodology section. For fine-tuning, we perform model parallelism-based distributed training on 4 A100 GPUs for 10 epochs with a learning rate of 5e-6 for spatial audio encoder, 1e-4 for Qformer projection and 2e-5 for Qwen3 4B thinking LLM. We have cross-entropy as the loss function. We found that a batch size of 8 per device work best. The low-rank adaptors (LoRA) hyperparameters alpha and r are set to 16 and 8, respectively. 

\section{Results}

Across all evaluations, we observe a consistent interaction between masking quality and thinking. We refer to the configuration without source separation as NoMask. While NoMask provides the lowest performance and GT masking provides an upper bound to the performance gains from time-based masking, AGM offers a middle ground, improving accuracy over NoMask and enabling some of the benefits associated with reasoning. From Table \ref{tab:by-qtype-grouped}, BAT baseline provides poor performance due to overfitting to it's training format questions which are mostly predicting the spatial attributes with limited reasoning. In the next subsections, we discuss and analyze our models' performance across various categories.

\begin{table*}[t]
\centering
\large
\setlength{\tabcolsep}{4pt}
\caption{Accuracy (\,\%). 
$\Delta\text{Interaction}$ measures how much more (or less) thinking helps under a method compared to NoMask:
$\Delta\text{Interaction} =
(\mathrm{Acc}_{T}^{\text{method}} - \mathrm{Acc}_{N}^{\text{method}})
-
(\mathrm{Acc}_{T}^{\text{NoMask}} - \mathrm{Acc}_{N}^{\text{NoMask}})$.
Acc is accuracy, T refers to thinking and N refers to non-thinking}
\label{tab:by-qtype-grouped}
\begin{tabular}{l c ccc ccc cc}
\toprule
 & \textbf{BAT} 
 & \multicolumn{3}{c}{\textbf{Thinking}} 
 & \multicolumn{3}{c}{\textbf{Non-Thinking}} 
 & \multicolumn{2}{c}{$\boldsymbol{\Delta}$ \textbf{Interaction (pp)}} \\
\cmidrule(lr){3-5}\cmidrule(lr){6-8}\cmidrule(lr){9-10}
\textbf{Question Type} 
 &  & NoMask & AGM & GT 
 & NoMask & AGM & GT 
 & AGM & GT \\
\midrule
Yes\_No          
 & 41.7 
 & 72.1 & 73.8 & \textbf{76.3}
 & 72.6 & 72.8 & 72.3 
 & \textbf{+1.5} & \textbf{+4.5} \\

Multiple Choice  
 & 0.3
 & \textbf{53.1} & 52.9 & 52.8
 & 50.4 & 50.3 & 50.3 
 & $-$0.1 & $-$0.2 \\

Open             
 & 0.00 
 & 41.0 & 41.4 & \textbf{41.8}
 & 41.2 & 41.0 & 41.5 
 & \textbf{+0.6} & \textbf{+0.5} \\
\bottomrule
  Overall 
  & 14.7
  & 54.3 & 55.0 & \textbf{56.1}
  & 54.1 & 54.0 & 54.1 
  & +0.8 & +1.8\\
\bottomrule
\end{tabular}
\end{table*}

\paragraph{Overall results:}
The Table \ref{tab:by-qtype-grouped} showcases the overall accuracy of the model. Overall, ground‑truth masking yields the highest accuracy (56.1\%), followed by AGM (55.0\%) and NoMask (54.3\%). Thinking improves performance most when masking is present: the improvement under GT masking is +2.0\% compared to only +0.2\% under NoMask, resulting in a  $\Delta$ Interaction of +1.8 pp. AGM shows an improvement of +1.0\% in between NoMask and GT.
 This improvement aligns with the intuition that the model benefits from both (i) cleaner evidence and (ii) an extended reasoning chain prior to generating the final answer.

\paragraph{Overall results by question type:} Across the three major question formats, Yes/No, Multiple‑Choice, and Open‑Ended, we observe clear differences in how source separation interacts with the LLM’s reasoning pipeline as seen in Table \ref{tab:by-qtype-grouped}.
For Yes/No questions, GT masking with thinking significantly improves accuracy. These questions tend to hinge on detecting a single dominant acoustic cue (“Does the Train whistle move from the right toward the left?”, “Is the crowing approaching the listener over time?”). Masking isolates the relevant source and removes competing background audio, making the evidence sharper and allowing the model to commit to a more confident binary decision.

For MCQ questions, AGM and GT masking perform on par or slightly lower than NoMask. We believe this could be due to two reasons: First, many MCQ questions require reasoning over multiple/all audio events. For example, “Which source spans a wider left->right range? choices: [Church bell, Electric shaver, They span equally wide, Can't tell]". Second, by only performing time-based masking, there is still some noise left in the audio clip from the irrelevant audio events hindering the LLM reasoning. In the future, we will work on performing time-frequency masking.

For open‑ended questions, we see a slight increase in accuracy for AGM and GT in thinking mode. These questions rely on a richer holistic audio scene (e.g., “Explain where each source ends and how its final position follows from its path.”, “Do the chainsaw and tick draw closer together, stay the same distance from each other, or move apart as the scene progresses? Explain perceptually.”).

\paragraph{Thinking over non-thinking:}
The $\Delta$ interaction measures how much thinking improves the performance of AGM or GT relative to NoMask. In Tables \ref{tab:by-qtype-grouped}, \ref{tab:xtab-qtype-overlap}, and \ref{tab:xtab-qtype-audio}, AGM delivers moderate gains in $\Delta$Interaction, while GT masking provides substantial gains, showing that high‑quality audio isolation most effectively supports text‑based reasoning. Overall, these results demonstrate that thinking mode reliably outperforms non‑thinking mode, and that its effectiveness is closely tied to the quality of source separation. Better‑isolated audio allows the LLM’s reasoning steps to operate on cleaner spatial evidence, resulting in stronger spatial AQA performance.

For NoMask, thinking provides only small improvements for Yes/No and Open‑Ended questions, with a slightly larger gain for Multiple‑Choice questions (+2.7\%). However, we also find that thinking incurs a notable latency cost: 6.19 seconds on average compared to 2.36 seconds for non‑thinking inference (2.6× slower). Taken together, these results suggest that when masking is unavailable, the modest accuracy gains offered by Thinking mode may not justify the substantially higher inference time. 

\paragraph{Accuracy by overlap:}
The Table~\ref{tab:by-overlap-thinking-only} highlights that even in thinking mode, filtering relevant audio sources continues to provide measurable benefits. For non overlapping audio event cases, GT masking yields highest gains of +2.9\% while AGM provides moderate yeild of +1.4\% over NoMask. In overlapping scenes, performance decreases across all methods, but the relative ordering of \emph{NoMask < AGM < GT} remains the same, indicating that reasoning is still limited by the residual noise left in the audio by irrelevant overlapped audio events after time-based masking.

\begin{table}[t]
\centering
\large
\setlength{\tabcolsep}{6pt}
\caption{Accuracy by overlap category (\%) for thinking mode.}
\label{tab:by-overlap-thinking-only}
\begin{tabular}{l ccc}
\toprule
\cmidrule(lr){2-4}
\textbf{Overlap} & NoMask & AGM & GT \\
\midrule
No overlap & 55.0 & 56.4 & \textbf{57.9} \\
Overlap    & 53.7 & 53.7 & \textbf{54.3} \\
\bottomrule
\end{tabular}
\end{table}


\begin{table}[t]
\centering
\large
\setlength{\tabcolsep}{6pt}
\caption{Accuracy by audio events (\%) for thinking mode.}
\label{tab:by-audio-thinking-only}
\begin{tabular}{l ccc}
\toprule
\cmidrule(lr){2-4}
\textbf{Audio Events} & NoMask & AGM & GT \\
\midrule
Complete match & 48.4 & 48.6 & \textbf{49.2} \\
1 Event        & 64.3 & 66.2 & \textbf{68.7} \\
2 events       & \textbf{54.1} & 53.7 & 53.8 \\
\bottomrule
\end{tabular}
\end{table}

\begin{table}[t]
\centering
\normalsize
\begin{tabular}{p{4cm} p{1.5cm}p{1cm}}
\toprule
\textbf{Error Type} & \textbf{Non-Thinking} & \textbf{Thinking} \\
\midrule
Lateral Direction        & 18.5 & 15.4 \\
Attribute Swapping       & 15.7 & 14.4 \\
Source Misidentification & 14.5 & 13.3 \\
Distance / Depth         & 14.2 & 10.3 \\
Crossing Center          & 11.7 & 8.9  \\
Temporal Sequence        & 7.8  & 5.5  \\
Rationale–Answer Mismatch & 9.5 & 4.1 \\
Magnitude Errors         & 2.9  & 2.7 \\
\bottomrule
\end{tabular}
\caption{Error category comparison for GT masking: Thinking vs Non-Thinking. Values are in \%. The lower error percentage the better.}
\label{tab:gt-masking-errors}
\end{table}

\paragraph{Accuracy by number of relevant audio events:}

The Table~\ref{tab:by-audio-thinking-only} summarizes performance across scenes containing a single event, two events, or all events (\texttt{Complete match}) with thinking mode. The \texttt{1 Event} condition yields the strongest improvements: GT+Thinking reaches 68.7\% with a +4.4\% gain over NoMask, and AGM also provides meaningful benefits (66.2\%, +1.9\%). This indicates that clean source isolation is crucial for leveraging the thinking mode. A similar but weaker trend appears in the \texttt{Complete match} setting, where both AGM and GT offer only modest performance gains, suggesting that mixed scenes with multiple active events are inherently harder to disentangle. Overall, the reduced gains for \texttt{Complete match} and \texttt{2 Events} highlight that multi‑source spatial mixtures remain a key bottleneck for reasoning even when masking is applied.

\paragraph{Comparison of error categories:}
Table~\ref{tab:gt-masking-errors} shows the categorization of errors in GT‑masked thinking and non‑thinking responses. GT masking in thinking mode reduces errors consistently across all categories, with the largest drops occurring in lateral direction (azimuth) errors (–3.1\%), distance/depth estimation errors (–3.9\%), and rationale–answer mismatches (–5.4\%). These improvements suggest that masking meaningfully enhances the precision of spatial geometry reasoning. Both azimuth and distance estimation rely heavily on clean directional cues, which are often blurred by overlapping sources in the non‑masked audio; isolating the ground‑truth region provides clearer spatial evidence and directly reduces these errors. The reduction in rationale–answer mismatches further indicates that thinking mode, when combined with masking, produces more stable reasoning and is less likely to generate rationales that contradict or diverge from the final answer. This aligns with the hypothesis that clearer audio inputs lead to more coherent and consistent reasoning trajectories.

\section{Conclusion}
In this work, we developed a spatial audio multimodal LLM capable of reasoning about dynamic audio scenes and tracking the movement of sound sources over time. Our architecture integrates spatial audio encoding, audio–language alignment, and LLM-based reasoning to support richer spatial understanding than prior static spatial AQA systems. A key finding is that incorporating query‑relevant source separation as a preprocessing filter meaningfully improves attention and downstream reasoning by isolating the most informative regions of the audio clip. To support progress in this area, we also introduced a movement‑focused spatial audio QA benchmark, enabling systematic evaluation of spatial motion understanding. Looking forward, we aim to extend our framework to more complex real‑world environments and multi-source recordings, and to explore stronger temporal and motion priors for spatial reasoning for long audio. 


\section*{Limitations}
One of the limitations in our pipeline is with the masking of the relevant audio events. Since we apply time based masking, the overlapping irrelevant events won't be suppressed completely. 

\bibliography{custom}

\appendix

\onecolumn  
\section{Prompt for Spatial Audio QA Pair Generation (with Thinking Steps)}
\label{app:spatial-audio-qa-prompt}
The following is the full prompt used to generate spatial audio question--answer pairs with perceptual thinking steps.
\begin{verbatim}
You are an expert in spatial audio reasoning and question generation. Your task is to create
high-quality QA pairs that evaluate a multimodal LLM’s understanding of spatial audio scenes.

### Inputs:
- FrameTrends: A dictionary where each key is an event name and each value contains:
    - azimuth: {direction, span_category, crosses_center, start_side, end_side, is_arc}
    - distance: {trend_profile, variation_category}
    - temporal: {start_time, end_time, duration} if available
    - summary_text: A natural-language summary combining lateral, radial, and temporal info
    - Note: direction may include 'arc left->right' or 'arc right->left' for curved trajectories

This FrameTrends object is derived from all available scene data, 
so treat it as the authoritative source for reasoning.

### Objectives:
Create diverse QA pairs that require reasoning about:
- Lateral trajectories (left->right, right->left, static)
- Arc trajectories (curved motion around the listener with constant radius)
- Radial changes (approach, recede, approach->recede)
- Relative motion between events (opposite directions, convergence/divergence)
- Sequencing and choreography (order, endpoints, overlap vs non-overlap if available)
- Comparative diagnostics (which spans wider azimuth? which ends farther?)
- Natural perceptual implications (e.g., “Which sound feels most noticeable near the middle?”)

### Constraints:
- DO NOT ask for timestamps or exact numeric values.
- Use only qualitative info from FrameTrends (direction, span_category, trend_profile, overlaps).
- Use listener-centric, natural language: “left”, “right”, “middle”, “closer”, “farther”.
- Avoid technical jargon like “binaural salience”; use intuitive terms like
“noticeable” or “prominent”.
- Everything occurs within -90° to +90° in front of the listener, 
so do not repeat “front stage” unnecessarily.
- Center crossing is optional; only mention if relevant.
- Avoid speculation beyond provided data; if ambiguous, 
use cautious phrasing (“appears to”, “likely”) only when supported.
- **Thinking steps and rationale must sound like perceptual reasoning, not metadata citation.**
    - Do NOT mention FrameData, MetaInfo, or structured sources.
    - Use natural language based on the scene description and qualitative trends 
    (e.g., “moves left to right and gets closer”).
    - Avoid phrases like “according to data”, “based on FrameTrends.” or "described as"
    - Write as if you are analyzing the raw audio features present in the embedding.

### Output Format:
Return an array of QA objects in JSON:
{
  "id": "<unique_id>",
  "question": "<string>",
  "type": "yes_no | multiple_choice | open",
  "choices": [optional for multiple_choice],
  "answer": "<string>",
   "thinking": "<think>\nStep 1: [Describe step]\nSolution: 
   [Provide scene-specific reasoning for this step]\n[Optional Step 2: ...]\nSolution: [...]\n[Optional Step 3: ...]
   \nSolution: [...]\nFinal reasoning: [Summarize why the answer is correct]\n</think>", 
  "rationale": "<one-sentence perceptual reasoning>",
  "tags": ["lateral", "radial", "relative_motion", "sequencing", "comparative", ...],
}

### Dynamic Thinking Steps Rules:
- For simple questions (single attribute, , one event, yes/no):
    - Use 1 step + solution +Final reasoning.
- For moderate questions (two attributes OR involves more than one event OR simple comparison):
    - Use 2 steps  + solutions + Final reasoning.
- For complex questions 
(comparative, multi-hop, choreography, relative motion, temporal reasoning):
    - Use 3 steps + solutions + Final reasoning.
- Always include Final reasoning summarizing why the answer is correct.
- Steps must be short, perceptual, and qualitative and followed by reasoning solution for that step 
(e.g., “Check if the sound moves laterally”).
- Do NOT include numeric values or cite structured data.

### Example Thinking Sections:
**Simple Yes/No Question**
<think>
Step 1: Check if the smoke detector moves laterally.
Solution: The smoke detector stays fixed on the right without sweeping.
Final reasoning: It remains on the right, so it is static in azimuth.
</think>

**Moderate Question (two events)**
<think>
Step 1: Compare lateral spans.
Solution: The train sweeps widely; the detector stays fixed.
Step 2: Compare radial behavior.
Solution: The train approaches then recedes; the detector remains steady.
Final reasoning: The train is dynamic both laterally and radially, 
while the detector is static.
</think>

**Complex Question**
<think>
Step 1: Identify lateral behavior of both sources.
Solution: The smoke detector is static on the right; the train moves right→left across the center.
Step 2: Identify radial behavior of both sources.
Solution: The smoke detector maintains a steady distance; the train approaches then recedes.
Step 3: Contrast them to explain differences.
Solution: One is stationary and steady; the other sweeps widely and varies strongly in distance.
Final reasoning: The train is dynamic both laterally and radially, while the detector is static.
</think>

### Required Mix:
- At least 5 Yes/No questions (fundamentals)
- At least 3 Multiple Choice questions (comparatives/diagnostics)
- At least 6 Open-ended questions (expanded reasoning), 
including at least 3 focused on RELATIVE MOTION between events
- At least 4 questions must involve radial reasoning (approach/recede)
- At least 2 questions should summarize overall choreography or sequencing
- At least 2 questions should compare lateral and radial trends together
- If temporal info exists, include at least 2 questions about order or overlap

### Open-Ended Reasoning Templates (adapt naturally):
1) **Qualitative Trend Summary**  
   - “Summarize how [Event] changes in distance while moving laterally.”
2) **Relative Motion Contrast**  
   - “Compare the movement of [Event A] and [Event B] in both direction and distance.”
3) **Stage Choreography**  
   - “Narrate the overall motion: where each source starts, how it moves, and where it finishes.”
4) **Convergence/Divergence**  
   - “Do [Event A] and [Event B] move toward the same region or away from each other? Explain.”
5) **Endpoint Reasoning**  
   - “Explain where each source ends and how its final position follows from its path.”
6) **Prominence Near Middle**  
   - “Which source feels most noticeable near the middle and why?”
7) **Comparative Span**  
   - “Which source spans a wider left–right range, and how does that affect the scene’s balance?”
8) **Temporal Order or Overlap**  
   - “Which event starts first?” or “Do [Event A] and [Event B] overlap in time?”
9) **Path Characteristics**  
   - "Describe the path that [Event] takes as it moves. 
   How does its distance change relate to its lateral movement?"
10) **Motion Pattern Comparison**  
   - "Compare how [Event A] and [Event B] move through space. 
   What makes their paths different?"

### Procedure (internal; do not output steps):
1) Use FrameTrends.summary_text and structured fields for reasoning.
2) Decide question complexity and adjust number of thinking steps dynamically:
    - If question involves >1 event -> at least moderate (minimum 2 steps).
3) For each step, provide both the step description and the solution 
for that step based on the scene.
4) Generate <think> section first, then answer, then rationale.
5) Ensure coverage of lateral, radial, relative motion, sequencing, and comparative reasoning.
6) Validate for internal consistency; if ambiguity exists, use cautious wording.
\end{verbatim}

\section{Evaluation of QA pairs}
\label{app:spatial-audio-qa-evaluation-prompt}
\subsection{Prompt for Evaluating Thinking Steps}
\noindent
\begin{verbatim}
       "You are evaluating the reasoning process (thinking) for a spatial audio question.

Question: {question}
Ground Truth Answer: {ground_truth_answer}

Ground Truth Thinking:
{ground_truth_thinking}

Predicted Thinking:
{predicted_thinking}

Evaluate the predicted thinking on four dimensions:

1. **Logical Coherence** (0-5): 
Does the reasoning flow logically from step to step?
   - 5: Perfect logical flow, each step follows naturally
   - 4: Mostly coherent with minor logical gaps
   - 3: Generally logical but some unclear connections
   - 2: Multiple logical gaps or unclear reasoning
   - 1: Poor logical structure
   - 0: Incoherent or contradictory reasoning

2. **Step Completeness** (0-5): 
Are all necessary reasoning steps present?
   - 5: All necessary steps present and well-developed
   - 4: Most steps present, minor omissions
   - 3: Key steps present but some missing
   - 2: Several important steps missing
   - 1: Many critical steps missing
   - 0: Minimal or no reasoning steps

3. **Factual Accuracy** (0-5):
Are the spatial facts and observations correct?
   - 5: All spatial facts correct (directions, distances, movements)
   - 4: Minor factual errors that don't affect conclusion
   - 3: Some factual errors but core understanding present
   - 2: Multiple factual errors affecting reasoning
   - 1: Major factual errors throughout
   - 0: Completely incorrect facts

4. **Alignment with Ground Truth** (0-5): 
How similar is the reasoning approach to the ground truth?
   - 5: Nearly identical reasoning approach and steps
   - 4: Very similar approach with minor differences
   - 3: Similar overall approach but different execution
   - 2: Different approach but reaches similar conclusions
   - 1: Significantly different approach
   - 0: Completely different reasoning

Also identify any specific errors in the thinking process. Categorize each error as:
- **Spatial reasoning errors**: lateral_direction, distance_depth, crossing_center,
temporal_sequence
- **Logical errors**: missing_steps, incorrect_flow, contradictions, incomplete_analysis
- **Factual errors**: source_misidentification, attribute_swapping, magnitude_errors
- **Consistency errors**: thinking_answer_mismatch, internal_contradictions

For each error, specify severity: minor, moderate, or major.

Provide your evaluation in this exact JSON format:
{{
    "logical_coherence": <score 0-5>,
    "step_completeness": <score 0-5>,
    "factual_accuracy": <score 0-5>,
    "alignment_with_ground_truth": <score 0-5>,
    "explanation": "<brief explanation of scores>",
    "errors": [
        {{"type": "<error_type>", "category": "<category>", "severity": "<severity>", 
        "description": "<description>"}},
        ...
    ]
}}"
\end{verbatim}

\subsection{Prompt for Evaluating Rationale}
\begin{verbatim}
"You are evaluating the rationale (brief explanation) for a spatial audio answer.

Question: {question}
Ground Truth Answer: {ground_truth_answer}
Predicted Answer: {predicted_answer}

Ground Truth Rationale:
{ground_truth_rationale}

Predicted Rationale:
{predicted_rationale}

Evaluate the predicted rationale on three dimensions:

1. **Conciseness** (0-5): Is the rationale appropriately brief yet complete?
   - 5: Perfect balance of brevity and completeness
   - 4: Mostly concise with minor verbosity or slight incompleteness
   - 3: Acceptable length but could be more concise or complete
   - 2: Too verbose or too brief, missing key points
   - 1: Significantly too long or too short
   - 0: Extremely verbose or minimal content

2. **Accuracy** (0-5): Does it correctly summarize the spatial relationships?
   - 5: All spatial facts correct and well-summarized
   - 4: Minor inaccuracies that don't affect understanding
   - 3: Some inaccuracies but core message correct
   - 2: Multiple inaccuracies affecting the summary
   - 1: Major inaccuracies throughout
   - 0: Completely incorrect summary

3. **Clarity** (0-5): Is it easy to understand?
   - 5: Crystal clear and easy to understand
   - 4: Clear with minor ambiguities
   - 3: Generally clear but some confusing parts
   - 2: Somewhat unclear or confusing
   - 1: Very unclear or hard to follow
   - 0: Incomprehensible

Also identify any specific errors in the rationale. Categorize each error as:
- **Spatial reasoning errors**: lateral_direction, distance_depth, crossing_center, 
temporal_sequence
- **Factual errors**: source_misidentification, attribute_swapping, magnitude_errors
- **Consistency errors**: rationale_answer_mismatch, contradictions

For each error, specify severity: minor, moderate, or major.

Provide your evaluation in this exact JSON format:
{{
    "conciseness": <score 0-5>,
    "accuracy": <score 0-5>,
    "clarity": <score 0-5>,
    "explanation": "<brief explanation of scores>",
    "errors": [
        {{"type": "<error_type>", "category": "<category>", "severity": "<severity>", 
        "description": "<description>"}},
        ...
    ]
}}"    
\end{verbatim}

\subsection{Evaluation of Open ended answers}
\begin{verbatim}
    "You are evaluating a spatial audio reasoning answer. 
    Compare the predicted answer against the ground truth.

Question: {question}

Ground Truth Answer: {ground_truth}

Predicted Answer: {predicted}

Evaluate the predicted answer on two dimensions:

1. **Factual Accuracy** (0-5): Are the spatial facts correct (directions, distances, movements, 
source identifications)?
   - 5: All facts are correct
   - 4: Minor factual errors that don't change the main conclusion
   - 3: Some factual errors but core understanding is present
   - 2: Multiple factual errors affecting the conclusion
   - 1: Major factual errors throughout
   - 0: Completely incorrect facts

2. **Semantic Similarity** (0-5): How well does the predicted answer capture the meaning and 
completeness of the ground truth?
   - 5: Captures all key points with equivalent meaning
   - 4: Captures most key points with minor omissions
   - 3: Captures main idea but misses some important details
   - 2: Partially captures the idea with significant gaps
   - 1: Minimal semantic overlap
   - 0: Completely different meaning

Provide your evaluation in this exact JSON format:
{{
    "factual_accuracy": <score 0-5>,
    "semantic_similarity": <score 0-5>,
    "explanation": "<brief explanation of scores>"
}}"
\end{verbatim}

\section{Further Results}
\subsection{Overall results}
\begin{table*}[t]
\centering
\large
\setlength{\tabcolsep}{4pt}
\caption{Overall core metrics. Accuracies are \%; Thinking/Rationale are 1--5 scales. GT refers to ground‑truth masking based source separation, NoMask refers to without using source separation, AGM refers to audio grounding model based source separation.}
\label{tab:overall-core-metrics}
\begin{tabular}{l ccc ccc cc}
\toprule
 & \multicolumn{3}{c}{\textbf{Thinking}} & \multicolumn{3}{c}{\textbf{Non-Thinking}}  \\
\cmidrule(lr){2-4}\cmidrule(lr){5-7}
\textbf{Metric} & NoMask & AGM & GT & NoMask & AGM & GT \\
\midrule
Accuracy (\%) & 54.3 & 55.0 & \textbf{56.1} & 54.1 & 54.0 & 54.1 \\
Thinking Avg (1-5) & 3.507 & 3.522 & \textbf{3.549} & 0.0 & 0.0 & 0.0 \\
Rationale Avg (1-5) & 4.156 & 4.168 & \textbf{4.179} & \textbf{4.182} & \textbf{4.182} & 4.180 \\
\bottomrule
\end{tabular}
\end{table*}

\subsection{Question Type x Audio Events}
\paragraph{Question type × audio events extracted:}
Table \ref{tab:xtab-qtype-overlap} shows the performance of models for each question type and relevant audio events frequency. For \texttt{Yes/No} and \texttt{Open} questions, the \texttt{1 Event} condition produces the highest accuracies across all masking methods. Both question types rely on identifying the salient motion of a single audio event either via GT masking or partially via AGM, reduces scene clutter and supports stronger reasoning gains. This is reflected in the large interaction values for \texttt{Yes/No} under 1 Event (+7.8 for GT, +2.6 for AGM) and moderate but positive gains for Open.
In contrast, the \texttt{Complete Match} condition offers smaller benefits from masking. MCQ questions often require 2 or more audio events to be compared across the spatial attributes. This reduces the necessity for masking in such scenarios. 
AGM shows this behavior most clearly as it helps in \texttt{Open} and \texttt{Yes/No} questions for 1 Event, but converges toward or below NoMask for MCQ in 1 Event and 2 Events conditions.
\begin{table*}[t]
\centering
\normalsize
\setlength{\tabcolsep}{3.8pt}
\caption{Question Type $\times$ Audio Events. 
Accuracies are \%; $\Delta$Interaction is defined as 
$(\mathrm{Acc}_T^{\text{method}} - \mathrm{Acc}_N^{\text{method}}) 
 - 
 (\mathrm{Acc}_T^{\text{NoMask}} - \mathrm{Acc}_N^{\text{NoMask}})$.}
\label{tab:xtab-qtype-audio}
\begin{tabular}{ll ccc ccc cc}
\toprule
 & & \multicolumn{3}{c}{\textbf{Thinking}} 
   & \multicolumn{3}{c}{\textbf{Non-Thinking}} 
   & \multicolumn{2}{c}{$\boldsymbol{\Delta}$\textbf{ Interaction (pp)}} \\
\cmidrule(lr){3-5}\cmidrule(lr){6-8}\cmidrule(lr){9-10}
\textbf{QType} & \textbf{Audio Events} 
  & NoMask & AGM & GT 
  & NoMask & AGM & GT 
  & AGM & GT \\
\midrule
Multiple Choice & Complete Match & \textbf{53.2} & 53.0 & 53.9 
                                          & 50.6 & 50.0 & 51.0 
                                          & \textbf{+0.4} & \textbf{+0.3} \\
Multiple Choice & 1 Event        & \textbf{57.0} & 56.3 & 52.7 
                                          & 52.7 & 54.2 & 50.2 
                                          & $-$2.2 & $-$1.8 \\
Multiple Choice & 2 Events       & \textbf{51.1} & 50.9 & 47.7 
                                          & 48.7 & 50.0 & 47.0 
                                          & $-$1.5 & $-$1.7 \\
Open            & Complete Match & \textbf{35.6} & 36.1 & 36.2 
                                          & 36.1 & 35.5 & 36.3 
                                          & \textbf{+1.1} & \textbf{+0.4} \\
Open            & 1 Event        & 52.8 & 53.2 & \textbf{54.2} 
                                          & 52.3 & 52.3 & 52.7 
                                          & \textbf{+0.4} & \textbf{+1.0} \\
Open            & 2 Events       & \textbf{43.1} & 43.0 & 43.6 
                                          & 43.3 & 43.7 & 43.2 
                                          & $-$0.5 & \textbf{+0.6} \\
Yes No          & Complete Match & 72.7 & 73.2 & \textbf{74.2} 
                                          & 72.4 & 72.4 & 72.8 
                                          & \textbf{+0.5} & \textbf{+1.1} \\
Yes No          & 1 Event        & 70.9 & 73.8 & \textbf{77.5} 
                                          & 72.5 & 72.8 & 71.3 
                                          & \textbf{+2.6} & \textbf{+7.8} \\
Yes No          & 2 Events       & \textbf{75.8} & 74.8 & 76.2 
                                          & 73.4 & 73.6 & 75.3 
                                          & $-$1.2 & $-$1.5 \\
\bottomrule
\end{tabular}
\end{table*}

\section{Question Type x Overlap}

\begin{table*}[t]
\centering
\normalsize
\setlength{\tabcolsep}{3.8pt}
\caption{Question Type $\times$ Overlap. Accuracies are \%; }
\label{tab:xtab-qtype-overlap}
\begin{tabular}{ll ccc ccc cc}
\toprule
 & & \multicolumn{3}{c}{\textbf{Thinking}} & \multicolumn{3}{c}{\textbf{Non-Thinking}} & \multicolumn{2}{c}{$\boldsymbol{\Delta}$ \textbf{Interaction (pp)}} \\
\cmidrule(lr){3-5}\cmidrule(lr){6-8}\cmidrule(lr){9-10}
\textbf{QType} & \textbf{Overlap} & NoMask & AGM & GT & NoMask & AGM & GT & AGM & GT \\
\midrule
Yes No          & No Overlap  & 72.2 & 74.4 & \textbf{78.3} & 72.6 & 73.0 & 72.2 & \textbf{+1.8} & \textbf{+6.5} \\
Yes No          & Overlap     & 72.0 & 73.2 & \textbf{74.4} & 72.6 & 72.6 & 72.4 & \textbf{+1.2} & \textbf{+2.6} \\
Multiple Choice & No Overlap & 53.2 & 55.3 & 54.2 & 49.2 & 47.6 & 48.4 & \textbf{+3.7} & \textbf{+1.8} \\
Multiple Choice & Overlap     & \textbf{53.0} & 50.5 & 51.4 & 51.6 & 53.0 & 52.2 & $-$3.9 & $-$2.2 \\
Open            & No Overlap  & 42.5 & 43.0 & \textbf{43.8} & 41.8 & 41.2 & 41.9 & \textbf{+1.1} & \textbf{+1.2} \\
Open            & Overlap     & 39.6 & 39.8 & \textbf{39.8} & 40.6 & 40.8 & 41.0 & \textbf{+0.0} & $-$0.2 \\
\bottomrule
\end{tabular}
\end{table*}

\section{Qualitative Evaluation}
Table\ref{tab:qualitative_eval} shows that AGM and GT masking substantially improve answer correctness across all question types. This improvement comes not only from clearer audio input but also from more accurate and grounded reasoning steps. In contrast, the no-mask conditions frequently exhibit direction flips, incorrect sweep-span assumptions, and mistakes in approach–recede interpretation even when detailed reasoning is provided. These patterns reveal that small ambiguities in audio can distort the entire reasoning chain, underscoring the importance of audio source separation. With disentangled audio, the model’s reasoning stabilizes and aligns with the true spatial cues, demonstrating that high-quality audio representations are essential for reliable audio-based reasoning. 

\begin{longtable}{p{0.27\linewidth} p{0.70\linewidth}}
\caption{Selected samples to illustrate model reasoning and responses across masking regimes}
\label{tab:qualitative_eval} \\
\toprule
\textbf{Question} & \textbf{Model Responses} \\
\midrule
\endfirsthead

\toprule
\textbf{Question} & \textbf{Model Responses (continued)} \\
\midrule
\endhead

\bottomrule
\endfoot


\textbf{Which sound spans a wider left–right range? Choices : [Whimper (dog), Tick-tock, They span the same range]} &
\textbf{Ground truth answer:} Whimper (dog)

\vspace{3.9em}
\textbf{NoMask without thinking mode} \\
& \textbf{Thinking:} \\
& \textbf{Answer:} \textcolor{red}{Tick-tock} \\
& \textbf{Rationale:} Tick-tock is described as a wide sweep while the dog's movement is moderate, so Tick-tock spans more left–right range.

\vspace{0.5em}
 \textbf{NoMask with thinking mode} \\
& \textbf{Thinking:} Step 1: Compare the span categories of both sounds.  
Solution: Both the whimper and tick-tock are described with a wide lateral span.  
Step 2: Decide which is wider based on those descriptions.  
Solution: Since both are labeled wide, they span the same wide left–right range.  
Final reasoning: Both sources have wide sweeps, so neither spans a wider left–right range than the other. \\
& \textbf{Answer:} \textcolor{red}{They span the same range} \\
& \textbf{Rationale:} Both the whimper and tick-tock are described as having wide lateral spans, so they cover the same wide left–right range.

\vspace{0.5em}
 \textbf{AGM with thinking} \\
& \textbf{Thinking:} Step 1: Compare the azimuth span of the two sounds.  
Solution: The whimper has a wide left→right sweep crossing the center, while the tick-tock is static on the right.  
Step 2: Decide which is wider.  
Solution: The whimper clearly spans a wider left–right range than the static tick-tock.  
Final reasoning: The whimper's wide sweep contrasts with the tick-tock's fixed position, so the whimper spans more azimuth. \\
& \textbf{Answer:} \textcolor{green}{Whimper (dog)} \\
& \textbf{Rationale:} The whimper performs a wide left→right sweep, whereas the tick-tock remains fixed on the right, so the whimper spans more azimuth.

\vspace{0.5em}
 \textbf{GT with thinking} \\
& \textbf{Thinking:} Step 1: Compare the lateral span of the two sounds.  
Solution: The whimper has a wide sweep across the scene, while the tick-tock is static on the right.  
Step 2: Conclude which spans more left–right range.  
Solution: The whimper clearly spans a wider range than the static tick-tock.  
Final reasoning: Because the whimper sweeps widely from left to right while the tick-tock stays fixed on the right, the whimper spans the wider range. \\
& \textbf{Answer:} \textcolor{green}{Whimper (dog)} \\
& \textbf{Rationale:} The whimper performs a wide left→right sweep whereas the tick-tock remains static on the right, so the whimper spans more azimuth. \\

\midrule


\textbf{Does the wind chime sweep all the way across the center to the left side? (Yes/No)} &
\textbf{Ground truth answer:} no

\vspace{2em}
\textbf{NoMask without thinking mode} \\
& \textbf{Thinking:} \\
& \textbf{Answer:} \textcolor{red}{Yes} \\
& \textbf{Rationale:} The wind chime follows an arc that crosses the center and ends on the left side.

\vspace{0.5em}
\textbf{NoMask with thinking mode} \\
& \textbf{Thinking:} Step 1: Check the wind chime's lateral path.  
Solution: The wind chime moves from the right toward the left and is described as crossing the center.  
Final reasoning: Because it travels right→left and crosses the center, it goes all the way across to the left side. \\
& \textbf{Answer:} \textcolor{red}{yes} \\
& \textbf{Rationale:} The wind chime moves from right to left and crosses the center, so it reaches the left side.

\vspace{0.5em}
\textbf{AGM with thinking mode} \\
& \textbf{Thinking:} Step 1: Check the lateral span of the wind chime.  
Solution: The wind chime moves from right to left and is described as a narrow sweep that does not cross center.  
Final reasoning: Because its sweep is narrow and does not cross the center, it does not reach the left side and thus does not sweep across the middle to the left. \\
& \textbf{Answer:} \textcolor{green}{no} \\
& \textbf{Rationale:} The wind chime has a narrow right→left motion that stays on the right and does not cross center.

\vspace{0.5em}
\textbf{gt\_thinking} \\
& \textbf{Thinking:} Step 1: Check the wind chime's lateral sweep and whether it crosses center.  
Solution: The wind chime moves from right to left but is described as a narrow sweep that stays on the right and does not cross center.  
Final reasoning: Because its sweep is narrow and it remains on the right side, it does not cross the center to reach the left. \\
& \textbf{Answer:} \textcolor{green}{No} \\
& \textbf{Rationale:} The wind chime's narrow right→left sweep remains on the right and does not cross the center. \\

\midrule


\textbf{Which source both moves right→left and shows a receding distance?  choices: [
        Power tool,
        Coin (dropping),
        Bicycle bell,
        Both Power tool and Coin (dropping)
      } &
\textbf{Ground truth answer:} Power tool

\vspace{1.5em}
\textbf{NoMask without thinking mode} \\
& \textbf{Thinking:} \\
& \textbf{Answer:} \textcolor{red}{Both Power tool and Coin (dropping)} \\
& \textbf{Rationale:} Both the power tool and the coin move right→left and are described as receding, unlike the bicycle which approaches.

\vspace{0.5em}
\textbf{NoMask with thinking mode} \\
& \textbf{Thinking:} Step 1: Check lateral directions for each source.  
Solution: Power tool and Coin both move right→left; Bicycle bell moves left→right.  
Step 2: Check radial trends for each source.  
Solution: Power tool recedes, Coin recedes, Bicycle bell approaches.  
Step 3: Combine lateral and radial.  
Solution: Both Power tool and Coin move right→left and recede, so they match the description.  
Final reasoning: Since both Power tool and Coin move right→left and are receding, they are the correct pair. \\
& \textbf{Answer:} \textcolor{red}{Both Power tool and Coin (dropping)} \\
& \textbf{Rationale:} Both the Power tool and Coin travel right→left and are described as receding, so they fit the condition.

\vspace{0.5em}
\textbf{AGM with thinking mode} \\
& \textbf{Thinking:} Step 1: Check lateral direction for each source.  
Solution: Both Power tool and Coin move right→left.  
Step 2: Check radial trends for each source.  
Solution: Power tool is receding; Coin is approaching.  
Final reasoning: Only the Power tool both moves right→left and recedes. \\
& \textbf{Answer:} \textcolor{green}{Power tool} \\
& \textbf{Rationale:} The Power tool moves right→left and recedes, while the Coin moves right→left but approaches.

\vspace{0.5em}
\textbf{GT with thinking mode} \\
& \textbf{Thinking:} Step 1: Check lateral directions for both candidates.  
Solution: Both the Power tool and the Coin move right→left.  
Step 2: Check radial trends for both candidates.  
Solution: The Power tool recedes while the Coin approaches.  
Final reasoning: Only the Power tool matches both right→left lateral motion and receding distance. \\
& \textbf{Answer:} \textcolor{green}{Power tool} \\
& \textbf{Rationale:} The Power tool is the only source that combines a right→left sweep with a receding distance trend. \\

\midrule


\textbf{Summarize how Oink changes in distance while it moves from left to right. (Open ended)} &
\textbf{Ground truth answer:} Oink moves left to right while approaching the listener.

\vspace{0.5em}
\textbf{NoMask without thinking mode} \\
& \textbf{Thinking:} \\
& \textbf{Answer:} Oink moves from left to right while \textcolor{red}{first approaching the listener and then receding}, so it feels closer as it nears the center and then moves away toward the right. \\
& \textbf{Rationale:} Oink's left→right path includes an approach then a recede, so it feels closer near the center and then farther as it continues right.

\vspace{0.5em}
\textbf{NoMask with thinking mode} \\
& \textbf{Thinking:} Step 1: Identify Oink's lateral motion.  
Solution: Oink moves left→right in a wide sweep that crosses the center.  
Step 2: Describe its radial trend during that sweep.  
Solution: Its distance follows an approach→recede pattern, so it gets closer as it nears the center and then moves away.  
Final reasoning: The combination of a wide left→right sweep and an approach→recede distance profile means Oink comes in toward the listener near the middle and then recedes toward the right. \\
& \textbf{Answer:} Oink moves from left to right while \textcolor{red}{first approaching the listener and then receding}, producing a strong sense of coming closer near the middle before moving away. \\
& \textbf{Rationale:} Oink's left→right sweep is paired with an approach→recede distance profile, so it draws nearer near the center and then moves away.

\vspace{0.5em}
\textbf{AGM with thinking mode} \\
& \textbf{Thinking:} Step 1: Describe Oink's lateral motion.  
Solution: Oink travels left→right across a wide arc and crosses the center.  
Step 2: Describe Oink's radial trend during that lateral motion.  
Solution: Its distance profile is an approach with large variation, so it gets closer as it moves.  
Final reasoning: Combining the wide left→right sweep with a clear approach trend explains how Oink moves closer while sweeping across the scene. \\
& \textbf{Answer:} \textcolor{green}{Oink moves from left to right while approaching the listener}; its distance decreases as it sweeps across, producing a clear approach during the left→right motion. \\
& \textbf{Rationale:} Oink performs a wide left→right sweep while approaching, so it becomes closer as it moves across the listener.

\vspace{0.5em}
\textbf{GT with thinking} \\
& \textbf{Thinking:} Step 1: Identify Oink's lateral movement.  
Solution: Oink moves left→right with a moderate sweep and crosses the center.  
Step 2: Identify Oink's radial trend while moving.  
Solution: Oink follows an approach trend, meaning it becomes closer over its path.  
Final reasoning: Together, Oink sweeps across the listener and approaches, so it grows closer as it travels from left to right. \\
& \textbf{Answer:} \textcolor{green}{Oink moves from left to right and steadily approaches the listener}, becoming noticeably closer as it travels across the scene. \\
& \textbf{Rationale:} Oink's left→right sweep combined with an approaching distance profile means it becomes closer as it moves across the listener. \\

\midrule


\textbf{Do Humming and Purr move in the same lateral direction? (Yes/No)} &
\textbf{Ground truth answer:} yes

\vspace{1.5em}
\textbf{NoMask without thinking mode} \\
& \textbf{Thinking:} \\
& \textbf{Answer:} \textcolor{red}{no} \\
& \textbf{Rationale:} Humming goes left→right while Purr goes right→left, so their lateral directions differ.

\vspace{0.5em}
\textbf{NoMask with thinking mode} \\
& \textbf{Thinking:} Step 1: Compare the lateral directions of both sounds.  
Solution: Humming moves left→right while Purr moves right→left, so they move in opposite directions.  
Final reasoning: Their lateral directions are opposite, not the same. \\
& \textbf{Answer:} \textcolor{red}{No} \\
& \textbf{Rationale:} Humming goes left→right while Purr goes right→left, so they move oppositely.

\vspace{0.5em}
\textbf{AGM with thinking mode} \\
& \textbf{Thinking:} Step 1: Compare the lateral directions of both sounds.  
Solution: Both Humming and Purr are described as moving left→right across the scene.  
Final reasoning: Since both are moving left→right, they share the same lateral direction. \\
& \textbf{Answer:} \textcolor{green}{yes} \\
& \textbf{Rationale:} Both Humming and Purr sweep from left to right, so their lateral directions match.

\vspace{0.5em}
\textbf{GT with thinking mode} \\
& \textbf{Thinking:} Step 1: Compare the lateral directions of Humming and Purr.  
Solution: Both are described as moving right→left.  
Final reasoning: Since both travel from right to left, they share the same lateral direction. \\
& \textbf{Answer:} \textcolor{green}{yes} \\
& \textbf{Rationale:} Both Humming and Purr sweep right→left, so their lateral directions are the same. \\

\end{longtable}

\end{document}